\documentclass{Interspeech2024}
\usepackage{multirow}
\usepackage{pifont}

\interspeechcameraready 

\title{Enhancing Partially Spoofed Audio Localization with Boundary-aware Attention Mechanism}

\name[affiliation={1}]{Jiafeng}{Zhong}
\name[affiliation={1*}]{Bin}{Li}
\name[affiliation={2}]{Jiangyan}{Yi}

\address{
  $^1$Guangdong Provincial Key Laboratory of Intelligent Information Processing, Shenzhen Key Laboratory of Media Security, Shenzhen University, China.
  $^2$State Key Laboratory of Multimodal Artificial Intelligence Systems, Institute of Automation,Chinese Academy of Sciences, China}
\email{zhongjiafeng2022@email.szu.edu.cn, libin@szu.edu.cn, jiangyan.yi@nlpr.ia.ac.cn}


\begin{document}

\maketitle

\begin{abstract}
   The task of partially spoofed audio localization aims to accurately determine audio authenticity at a frame level. Although some works have achieved encouraging results, utilizing boundary information within a single model remains an unexplored research topic. In this work, we propose a novel method called Boundary-aware Attention Mechanism (BAM). Specifically, it consists of two core modules: Boundary Enhancement and Boundary Frame-wise Attention. The former assembles the intra-frame and inter-frame information to extract discriminative boundary features that are subsequently used for boundary position detection and authenticity decision, while the latter leverages boundary prediction results to explicitly control the feature interaction between frames, which achieves effective discrimination between real and fake frames. Experimental results on PartialSpoof database demonstrate our proposed method achieves the best performance. The code is available at \url{https://github.com/media-sec-lab/BAM}.
\end{abstract}

\renewcommand{\thefootnote}{\fnsymbol{footnote}}
\footnote{Corresponding author.}
\setcounter{footnote}{0}
\renewcommand{\thefootnote}{\arabic{footnote}}

\section{Introduction}

With the rapid advancement of Artificial Intelligence Generated Content (AIGC), technologies like text-to-speech (TTS) \cite{shen2023naturalspeech} and voice conversion (VC) \cite{li2023freevc} can generate realistic human voices. Partially spoofed audio, where synthesized speech segments are inserted or spliced into genuine utterances, pose significant threats. An attacker can alter the meaning of sentences by manipulating small and specific units (e.g., words, characters, or even phonemes), deceiving both machines and humans.

In recent years, numerous methods and datasets have been developed for Partially Spoofed Audio Detection (PSAD). Yi \textit{et al.} \cite{yi21_interspeech} constructed the first partially spoofed audio dataset named HAD, which replaces some nature segments with synthesized speech segments having different semantic content. The Audio Deep Synthesis Detection (ADD) challenge 2022 \cite{yiadd2022} involves a detection track containing partially spoofed audio. The organizers provided only real and entirely fake data for training, and the task was target to detect fake at utterance level. In this challenge, notable detection performance was achieved by fine-tuning pretrained self-supervised learning (SSL) models with different back-end classifiers \cite{lvfakeaudio,liudeepspectro}. Boundary detection task was only introduced as a proxy task to achieve utterance-level detection \cite{caiadd2022, wu2022partially}.

Beyond merely detecting the presence of spoofing in audio, the Partially Spoofed Audio Localization (PSAL) task has become an emerging topic for better analyzing deepfake audio. Zhang \textit{et al.} \cite{zhang2021initial} developed a dataset named PartialSpoof, which includes segment-level labels. They then investigated an approach to detect spoofing at both segment and utterance levels within a multi-task learning framework \cite{zhang2021multi}. Subsequently, they extended the PartialSpoof dataset and used Wav2vec2 (W2V2) \cite{baevski2020wav2vec} as a front-end to detect segment-level and utterance-level fakes simultaneously \cite{zhang2022partialspoof}. Moreover, the ADD 2023 \cite{yi2023add} introduced a PSAL track, which has resulted in numerous promising works \cite{martin2023vicomtech,li2023multi,cai2024integrating}. Among them, an effective approach was proposed to achieve the best localization score by integrating the decisions from three countermeasure (CM) systems including boundary detection, frame-level fake detection, and utterance-level fake detection \cite{cai2024integrating}. Recently, an approach incorporating a contrastive learning  module and temporal convolution was proposed to effectively capture better features for localization \cite{xie2023efficient}.

\begin{figure}[t]
  \centering
  \includegraphics[width=\linewidth]{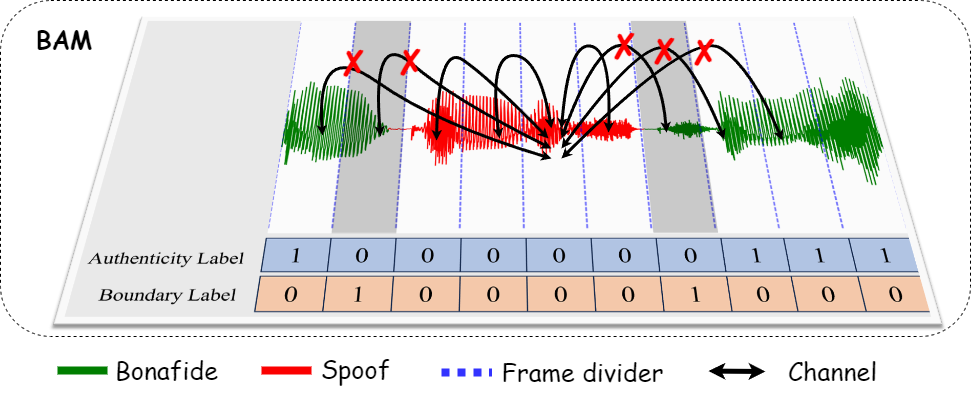}
  \caption{By introducing the frame-wise attention based on boundary prediction, BAM motivates each frame to exchange information with other frames with the same authenticity label.}
  \label{fig:introduction}
\end{figure}

Some aforementioned methods \cite{caiadd2022,wu2022partially,cai2024integrating} utilize boundary information solely for PSAD at utterance-level detection. However, leveraging boundary features to aid in pinpointing the exact locations of spoofing within a single CM system is still an unexplored topic. As we know, all frames within a segment delineated by boundaries are assigned the same label. Once the correct boundary position is identified, the authenticity of a frame can be determined by referencing other frames within the same segment. Thus, boundary information serves as vital auxiliary knowledge that can steer authenticity decisions at the frame level. To exploit this idea, in this paper, we have designed a new attention mechanism that leverages boundary information, as illustrated in Figure~\ref{fig:introduction}. Message passing between those frames that belong to different authenticity classes is explicitly weakened by the attention mechanism. Our proposed method is named Boundary-aware Attention Mechanism (BAM). Specifically, pre-trained W2V2 \cite{baevski2020wav2vec} or WavLM \cite{chen2022wavlm} is used as front-end feature extractor. Subsequently, a Boundary Enhancement (BE) module extracts inter-frame and intra-frame features to effectively detect the boundary frame position. Finally, a Boundary Frame-wise Attention (BFA) module leverages the boundary prediction results to make reliable  authenticity decisions at frame level. In this way, discriminative features are learned to better distinguish between real and fake frames, leading to improved localization performance. In summary, the contributions of this paper are as follows:
\begin{itemize}
\item We propose a novel approach called BAM that takes boundary information as auxiliary attentional hint to guide localization. To the best of our knowledge, this work is the first attempt to leverage boundary information for PSAL within a single CM.
\item We develop a boundary enhancement module containing two branches for better exploiting intra-frame and inter-frame information to obtain discriminative boundary features.
\item We conducted experiments to demonstrate the effectiveness of the proposed method. Compared with existing methods, our method achieves the best localization performance on PartialSpoof dataset.
\end{itemize}

\section{Proposed method}

\begin{figure*}[t]
  \centering
  \includegraphics[scale=0.5]{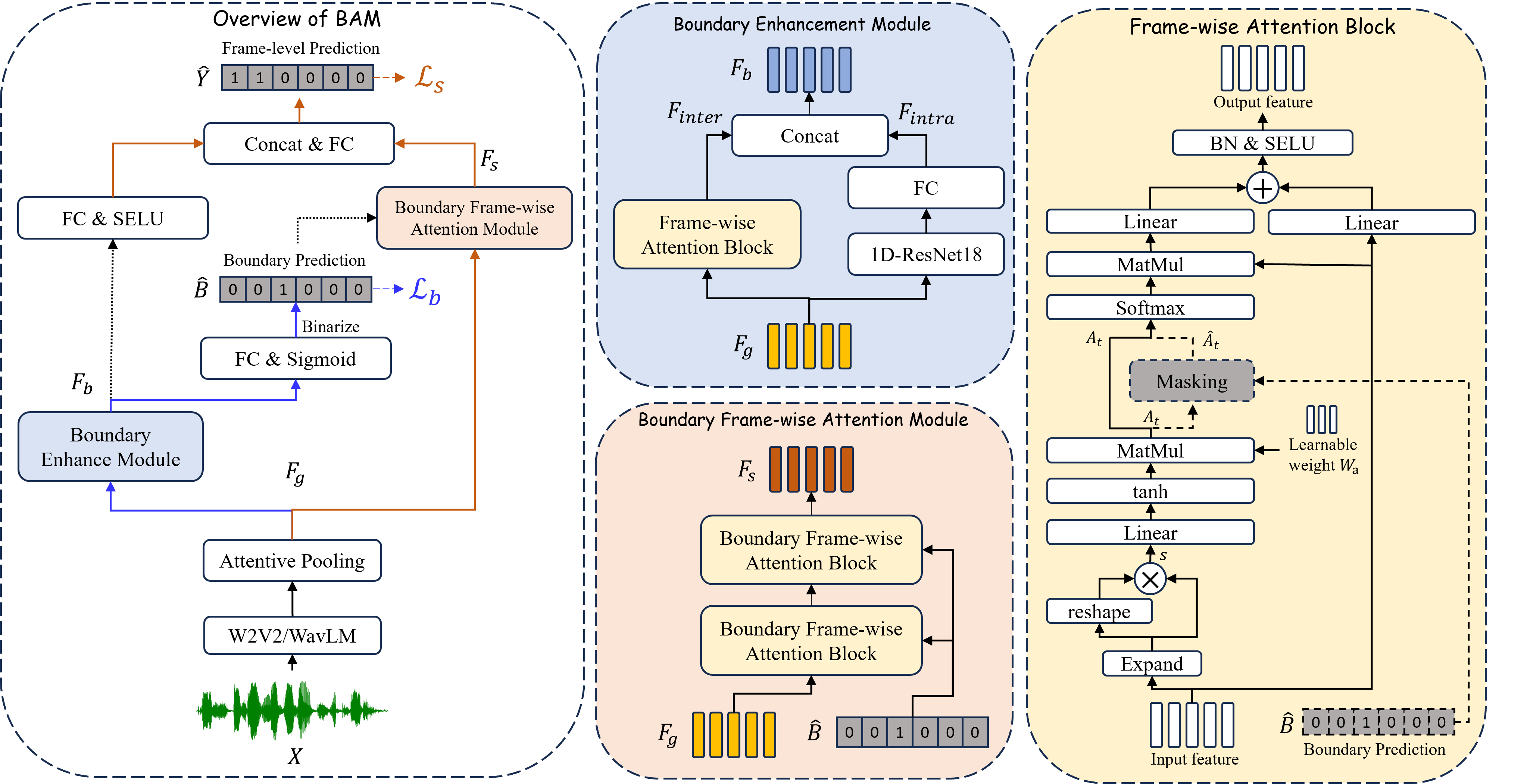}
  \caption{The architecture of our BAM. The BAM framework (left) comprises a Boundary Enhancement (BE) Module (center top) and a Boundary Frame-wise Attention (BFA) Module (center bottom). The dotted line arrows indicate no gradient propagation, while the solid arrows in different colors represent the gradient propagation corresponding to various loss functions. The Boundary Frame-wise Attention Block (BFAB) is identical to the Frame-wise Attention Block (FAB) but it is equipped with an additional boundary masking component (as indicated by dashed lines).}
  \label{fig:overview}
\end{figure*}

The flowchart of our BAM is depicted in Figure~\ref{fig:overview}. Firstly, we use a pre-trained SSL model to extract the features of speech and apply an attentive pooling layer \cite{okabe2018attentive} to make each frame represents a specific temporal resolution (e.g., \SI{160}{\milli\second}). The output of the pooling layer is fed into the Boundary Enhance (BE) module to enhance boundary feature representation and identify the boundary frames with a simple fully connected layer. Then, the Boundary-aware Frame-wise Attention (BFA) module takes the boundary prediction results and the outputs of the pooling layer as inputs to capture the correlation information among frames. Finally, the outputs of the BFA module and the BE module, are concatenated and fed into a fully-connected layer to make frame-level authenticity decisions.

\subsection{Pretrained self-supervised front-end}
We start by processing raw speech using a pretrained SSL speech model, W2V2 \cite{baevski2020wav2vec,babu2021xls} or WavLM \cite{chen2022wavlm}, to extract effective front-end features. Compared to traditional handcrafted features (e.g., LFCC \cite{todisco18_interspeech}, MFCC \cite{mfcc}), the SSL-based front-end features exploit extensive volumes of unlabeled data for pre-training, significantly improving data representation capabilities and facilitating the identification of intricate patterns within audio data. In our implementation, we utilize WavLM-large \cite{chen2022wavlm} or XLS-R-300M \cite{babu2021xls} and fine-tune their weights in conjunction with other modules, and use their last hidden states of the transformer as front-end features.

\subsection{Boundary enhancement module}

In the context of PSAL, frames at the boundaries that contain both spoofed and genuine samples are labeled as spoofed. During training with binary authenticity labels, boundary frames (especially those with a smaller proportion of spoofed data) are close to fully genuine frames in the feature space, even though they have opposite labels. This situation can lead to training instability and performance degradation. To mitigate this issue, we have developed a boundary enhancement module specifically to model and distinguish boundary frames from non-boundary frames.

As shown in Figure~\ref{fig:overview}, the boundary enhancement module transforms the front-end feature $F_g \in \mathbb{R}^{T \times D} $ into the boundary feature $F_b \in \mathbb{R}^{T \times 2D}$, where $F_g$ is the output feature of a pooling layer, $T$ is the number of frames, and $D$ is the feature dimension. The boundary feature serves dual purposes.. Firstly, it is fed into a fully-connected layer with a sigmoid function to obtain the boundary prediction probability $\hat{b} \in \mathbb{R}^{T \times 1}$ as following:
\begin{equation}
    \centering
    \label{eq:TAP}
    \hat{b}=sigmoid(\phi(F_b)), 
\end{equation}
where the linear mapping function $\phi$ with a learnable weight matrix $W$ and a bias term $b$ is given by: 
\begin{equation}
    \centering
    \label{eq:TAP}
    \phi(x)=Wx+b, 
\end{equation}
Then we conduct binarization with a fixed threshold to make binary boundary prediction $\hat{B}$. Secondly, the boundary feature is processed by a fully-connected layer with an activation function and concatenated for the final frame-level authenticity decision. 

The detail of BE module is illustrated in the middle of Figure~\ref{fig:overview}. Specifically, two different branches are designed to respectively extract the intra-frame and inter-frame features. For the inter-frame branch, a Frame-wise Attention Block (FAB) is employed to capture inter-frame correlations using an attention mechanism. The right part of Figure~\ref{fig:overview} illustrates the details of the FAB. We start by computing the attention score $s \in \mathbb{R}^{T \times T \times D}$ through the element-wise multiplication of each frame feature with others. Then, a learnable attention weight $W_a \in \mathbb{R}^{D\times H}$ is defined to weigh the attention score. This process can be written as:
\begin{equation}
    \centering
    \label{eq:at}
    A_{t}=tanh(\phi(s))W_a,
\end{equation}
where $A_{t} \in \mathbb{R}^{T\times T\times H}$ represents the attention map. Subsequently, each frame within $F_g$ aggregates information from other frames, utilizing the attention map as following:
\begin{equation}
    \centering
    \label{eq:TAP}
    F_{a}=softmax(A_{t})F_g,
\end{equation}
where $F_a \in \mathcal{R}^{H\times T\times D}$ represents updated frame features. The residual structure is employed to stabilize training. The output of FAB is obtained from $F_a$ and $F_g$ according to:
\begin{equation}
    \centering
    \label{eq:TAP}
    F_{inter}=\mathcal{S}(\mathcal{BN}((\phi(F_a) \oplus (\phi(F_g)))),
\end{equation}
where $\oplus$ represents element-wise addition and $\mathcal{BN}$ represents 1-D batch normalization, and $\mathcal{S}$ is SELU \cite{klambauer2017self} activation function. For the intra-frame branch, each frame is individually fed into a 1D-ResNet \cite{he2016deep}, followed by a fully-connected layer to learn intra-frame features. Then, the intra-frame feature $F_{intra}$ and inter-frame feature $F_{inter}$ are concatenated to obtain the boundary feature $F_b$. 

\subsection{Boundary frame-wise attention module}
\begin{figure}[t]
  \centering
  \includegraphics[width=\linewidth]{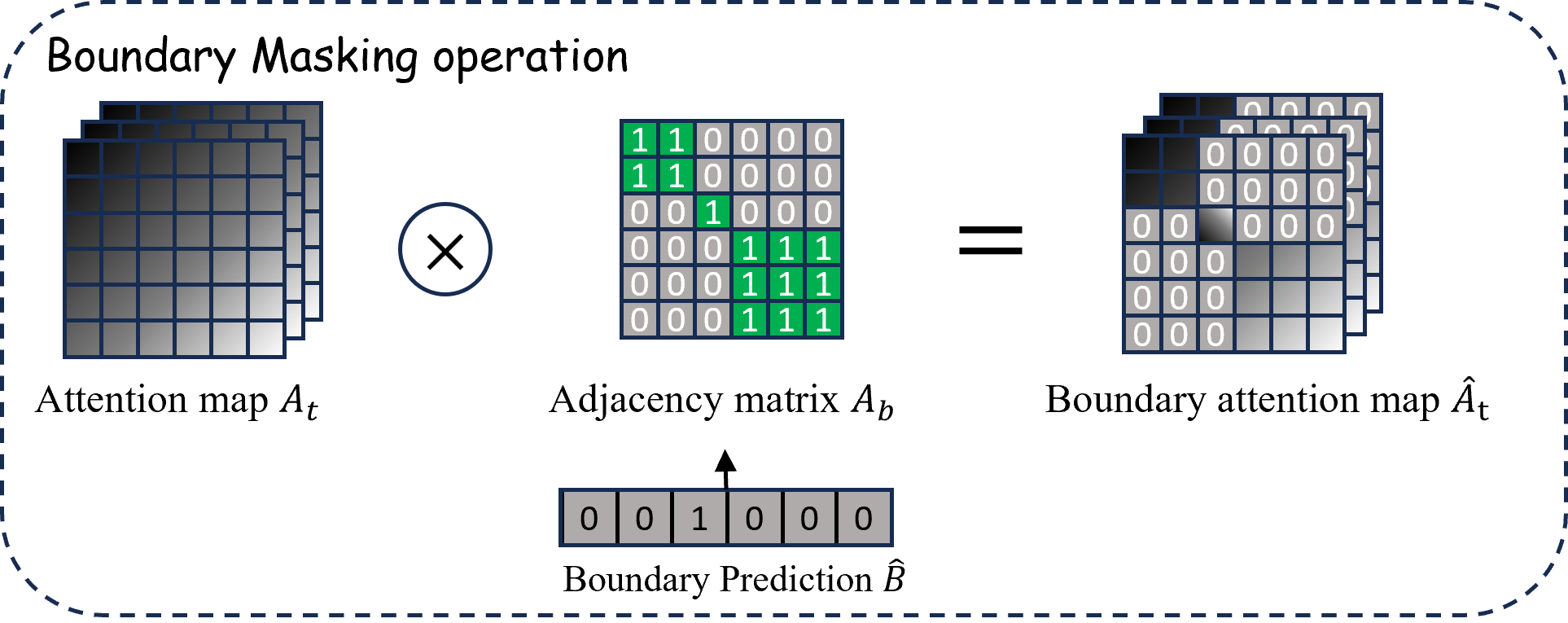}
  \caption{The illustration of the boundary masking operation. The $\otimes$ denotes element-wise multiplication.}
  \label{fig:boundary mask}
\end{figure}

As shown in Figure~\ref{fig:overview}, the BFAM module consists of two stacked Boundary Frame-wise Attention Blocks (BFAB). Specifically, the structure of BFAB is similar to FAB but is equipped with a boundary masking component, which takes binary boundary prediction $\hat{B}$ and $F_g$ to weaken the message passing between those frames that belong to different genres. A boundary adjacency matrix $A_{b}\in\{0,1\}^{T \times T}$ is constructed based on the boundary prediction $\hat{B}$. Its element is defined as follows:
\begin{equation}
    \centering
    \label{eq:mtrix}
    A_{{b}_{i,j}}=
    \begin{cases}
    1& \text{if } i = j, \\
    \prod_{n=i}^{j}{(1-\hat{B}[n])}& \text{if } i<j, \\
    \prod_{n=j}^{i}{(1-\hat{B}[n])}& \text{if } i>j,
    \end{cases}
\end{equation}
where $\hat{B}[n]$ represents the $n^{th}$ element of sequence $\hat{B}$, and $i,j\in[0,1,\dots,T-1]$ represent the indices of frames. In other words, equation~\ref{eq:mtrix} determines whether there is a boundary between the $i^{th}$ and the $j^{th}$ frame; if there is, the value in the $i^{th}$ row and the $j^{th}$ column of $A_{b}$ is set to 0, otherwise, it is set to 1. Once the adjacency matrix is obtained, the attention map (refer to equation~\ref{eq:at}) can be updated to a boundary attention map $\hat{A_{t}}$, as illustrated in Figure~\ref{fig:boundary mask}, and given as:
\begin{equation}
    \centering
    \label{eq:TAP}
    \hat{A_{t}}=A_{t} \otimes A_{b},
\end{equation}
where $\otimes$ denotes element-wise multiplication.

\subsection{Loss function}
We employ two loss functions to constrain the model for supervised training: boundary loss $\mathcal{L}_{b}$ and frame-level authenticity loss $\mathcal{L}_{s}$ as shown in Figure~\ref{fig:overview}. For boundary loss, binary cross-entropy loss is used and the ground-truth boundary labels $B$ can be derived from the segment-level authenticity labels $Y$. It is worth noting that we only set the boundary frames as label 1 and all other frames as label 0, which differs from previous methods \cite{caiadd2022,wu2022partially,cai2024integrating} where they also set the label of frames near the boundary frames to 1. This labeling strategy \cite{caiadd2022,wu2022partially,cai2024integrating} may introduce additional noise in our case and lead to diminished localization performance. For authenticity loss, we follow prior work and utilize the standard cross-entropy loss. In summary, the overall loss can expressed as:
\begin{equation}
    \centering
    \label{eq:TAP}
    L = \mathcal{L}_s(\hat{y},Y)+\lambda \mathcal{L}_b(\hat{b},B),  
\end{equation}
where $\hat{y} \in \mathbb{R}^{T\times 1}$ represents the authenticity prediction result of model, and $\lambda$ is set to 0.5 in practice.

\section{Experiments and results}

\subsection{Dataset and implementation details}
To validate the effectiveness of our method, we conducted experiments on the PartialSpoof \cite{zhang2022partialspoof}. The dataset is constructed by segmenting the fully spoofed and genuine utterances from the ASVspoof2019 \cite{todisco2019asvspoof} LA database using Voice Activity Detection (VAD) and then concatenating those segments with basic digital signal processing. Four evaluation metrics are used to compare model performance: Equal error rate (EER), precision, recall and $F_1$ score.   

In the training phase, for experiments conducted at a \SI{160}{\milli\second} temporal resolution, we fix the length of training samples at 4s. Samples exceeding this length are randomly trimmed, while those shorter than are padded with zero. Then raw data is padded to correspond with the \SI{20}{\milli\second} shift of W2V2 and WavLM. Therefore, the attentive pooling stride is set to 8, the number of frames $T$ is 25, the feature dimension $D$ is 1024, block number $N$ is 2, and the learnable weight head number $H$ is 1. For experiments conducted at a \SI{20}{\milli\second} temporal resolution, the number of frames $T$ is 200, and we use a linear transformation to reduce the feature dimension $D$ to 256. Other hyperparameters remain consistent with those used in \SI{160}{\milli\second} experiments. In the test phase, variable-length test samples are directly fed into the model to obtain the prediction result. we use Adam optimizer with a learning rate of $10^{-5}$ and halved every 10 epochs, and all of model are trained for 50 epochs.

\subsection{Comparison with existing methods}
To ensure a fair comparison with existing methods \cite{zhang2021initial,zhang2021multi,zhang2022partialspoof,cai2024integrating,xie2023efficient}, we conducted experiments with 160 ms temporal resolution. The results are shown in Table~\ref{tab:e1}. Based on our observation, WavLM achieves better results compared with W2V2-XLSR. Moreover, Our method gets EER as 3.58\% and F1 score as 96.09\%, reaching the best single CM performance on PartialSpoof dataset.

\begin{table}[t]
\centering
\caption{Comparison of EER (\%) and F1 score (\%) results with current methods and different front-end feature on PartialSpoof dataset. Results marked with * are those we have reproduced with the same setting, while the rest are from the original paper.}
\label{tab:e1}
\begin{tabular}{cccc} 
\hline
\bfseries Model   & \bfseries Front-end    & \bfseries EER   & \bfseries F1 score \\ 
\hline
LCNN-BLSTM \cite{zhang2021initial}   & LFCC       & 16.21 & -       \\ 
SELCNN-BLSTM \cite{zhang2021multi} & LFCC       & 15.93 & -       \\ 
Single reso. \cite{zhang2022partialspoof}  & W2V2-Large & 6.25  & -       \\ 
Multi reso. \cite{zhang2022partialspoof}  & W2V2-Large & 9.24  & -       \\ 
SPF \cite{cai2024integrating}          & W2V2-XLSR & - & 91.48           \\ 
SPF \cite{cai2024integrating}          & WavLM-Large  & - & 92.96               \\
TDL \cite{xie2023efficient}        &  W2V2-XLSR  & $10.98^{*}$ & $89.19^{*}$ \\
Ours        & W2V2-XLSR &  4.12 &    94.98     \\ 
Ours         & WavLM-Large & \textbf{3.58}  &  \textbf{96.09}     \\
\hline
\end{tabular}
\end{table}

\subsection{Ablation study}
\begin{table}[t]
\centering
\caption{The ablation experiment result (\%) for localization task.}
\label{tab:e2}
\begin{tabular}{cccccc} 
\hline
\bfseries Number & \bfseries Model  & \bfseries EER  & \bfseries F1 score & \bfseries Precision & \bfseries Recall   \\
\hline
\normalsize{\textcircled{\scriptsize{1}}}\normalsize&Baseline &  5.79  & 94.36 &  92.71  & 96.08 \\
\normalsize{\textcircled{\scriptsize{2}}}\normalsize&FA  & 4.62   & 95.78   & 93.55   & 98.11 \\
\normalsize{\textcircled{\scriptsize{3}}}\normalsize&FA+BE  &  3.91 & 96.01  & \textbf{93.78}   & 98.34 \\
\normalsize{\textcircled{\scriptsize{4}}}\normalsize&BFA+BE & \textbf{3.58} & \textbf{96.09} & 93.68 & \textbf{98.51} \\
\hline
\end{tabular}
\end{table}
We conducted ablation experiments to assess the effectiveness of each modules within our framework under two settings: 1) localization task and 2) boundary detection task. Table~\ref{tab:e2} presents the results for localization task. The baseline model, denoted as \normalsize{\textcircled{\scriptsize{1}} }\normalsize, is equipped with a WavLM front-end, succeeded by a max pooling layer and a fully-connected layer. The models \normalsize{\textcircled{\scriptsize{2}} }\normalsize and \normalsize{\textcircled{\scriptsize{3}} }\normalsize  represent simplified versions of the BAM. Model \normalsize{\textcircled{\scriptsize{2}} }\normalsize incorporates only the right branch as illustrated in Figure~\ref{fig:overview}, while model \normalsize{\textcircled{\scriptsize{3}} }\normalsize includes both the left and right branches but omits the middle pathway from the BE module to the BFA module. Model \normalsize{\textcircled{\scriptsize{4}} }\normalsize is actually the proposed BAM. It can be seen that each module contributes to the enhancement of localization performance. Specifically, compared to the baseline model, our method reduced the EER by 2.21\% and increased the F1 score by 1.73\%.

Table~\ref{tab:e3} presents the results for boundary detection task. It is clear that the inter-frame feature is more effective than intra-frame. The BE module combined with both inter-frame and intra-frame achieves the best boundary detection performance with EER as 3.33\% and F1 score as 92.25\%.
\begin{table}[t]
\centering
\caption{The ablation experiment result (\%) for boundary detection task.}
\label{tab:e3}
\begin{tabular}{ccccc} 
\hline
\bfseries Model  & \bfseries EER   & \bfseries F1 score & \bfseries Precision & \bfseries Recall \\
\hline
FC           & 4.50  &  89.93       &  90.19   & 89.67    \\
Inter-frame  & 3.59  &  91.77      &  91.46   &  92.07   \\
Intra-frame  & 3.79  &  91.61     & 91.63    & 91.59    \\
BE           & \textbf{3.33} & \textbf{92.25}     & \textbf{91.74}    & \textbf{92.78}    \\
\hline
\end{tabular}
\end{table}

\subsection{Finer-grained resolution experiment}

\begin{table}[t]
\centering
\caption{The finer-grained resolution experiment result (\%) for localization task. The model with a \SI{20}{\milli\second} resolution is trained from scratch as the base model, whereas the remaining models are finetuned from this base model.}
\label{tab:e4}
\begin{tabular}{cccc} 
\hline                            
\bfseries Resolution (ms)   & \bfseries EER   & \bfseries F1 score   \\                               
\hline
20        & 5.20  & 95.82          \\
40        & 4.90  & 95.90          \\
80         & 4.32  & \textbf{95.97}     \\
160          & 3.66  &  95.95     \\
320         & 2.71  &  95.88       \\
640         & \textbf{2.28}  &  95.64    \\
\hline
\end{tabular}
\end{table}

Extra experiment was conducted at a \SI{20}{\milli\second} temporal resolution. Specifically, we trained the BAM without pooling directly on the front-end feature with a \SI{20}{\milli\second} shift for 50 epochs as the base model. Subsequently, we removed the last fully-connected layer from pretrained base model and replaced it with an attentive pooling layer succeeded by a new fully-connected layer. We finetuned this model for 10 epochs to predict the authenticity at specific temporal resolutions. The results are shown in Table~\ref{tab:e4}. The base model achieves remarkable localization performance at \SI{20}{\milli\second} resolution with EER as 5.20\% and F1 score as 95.82\%. As the temporal resolution decreases, the F1 score remains almost constant while the EER continually increase. We explain the reason behind this trend is that with the decreases in resolution, the total number of frames and the proportion of genuine frames both rise, and EER is particularly sensitive to changes in class proportions while the F1 score is relatively robust. Moreover, there is a marginal decrease in localization performance compared to BAM executed on frames of \SI{160}{\milli\second} (3.66\% v.s 3.58\%). We attribute this decline in performance to the increased challenge of boundary detection at finer resolutions. The base model gets EER as 5.12\% and F1 score as 82.97\% at \SI{20}{\milli\second} resolution for boundary detection task, which is worse than the boundary detection performance at \SI{160}{\milli\second} (refer to Table~\ref{tab:e3}).
    
\section{Conclusion}
In this paper, we propose a novel partially spoofed audio localization method. Our method simultaneously conduct boundary detection and frame-level authenticity determination tasks within a single CM model. The boundary information is utilized to enhance the accuracy of localization. Experimental result show that the proposed method achieves the best performance on PartialSpoof dataset. The finer-grained resolution experiment demonstrates that accurately detecting boundary position at a finer resolution is a more challenging task and will be a direction of our future work.

\section{Acknowledgements}
This work was supported in part by the National Natural Science Foundation of China (Grant U23B2022, U22B2047, 62322120), and in part by the Guangdong Provincial Key Laboratory (Grant 2023B1212060076).

\bibliographystyle{IEEEtran}
\bibliography{mybib}

\end{document}